\documentclass[aps,prb,showpacs,reprint,amsmath,amssymb,superscriptaddress,a4paper]{revtex4-1}
\usepackage[utf8]{inputenc}
\usepackage{amsmath}
\usepackage{amsfonts}
\usepackage{amssymb}
\usepackage{graphicx}
\usepackage{upgreek}
\usepackage{xcolor}
\usepackage[english]{babel}
\usepackage{lipsum}
\usepackage{lineno}
\usepackage{parskip}
\begin{document}
%\title{Small footprint optoelectronic neuron with excitation, inhibition and non-linear activation}
%\title{Nanoscale optoelectronic neuron with excitation, inhibition and tunable activation}
%\title{Modular nanoscale photosensitive artifical neuron with excitation, inhibition and tunable activation}
%\title{Optoelectronic nanowire neuron for nanoscaled and modular neuromorphics}
\title{Nanoscale photonic neuron with biological signal processing}

%\title{Modular nanowire neuron with photoactivated and tunable excitation-inhibition function}

%with excitation, inhibition and tunable activation

%\title{Photosensitive nanowire neuron for modular neuromorphics with excitation, inhibition and tunable activation}

%\title{Optoelectronic nanowire neuron}

%%% Overall structural suggestions:
% ---- Format paper to fit a Nature letter (not an 'article')
% ---- Nature Letters formatting guidelines on Nature Letters are as follows: 
% ---- Introductory paragraph (not abstract) up to 150 words, summarizing the background, rationale, main results and implications. This paragraph should be referenced, and should be considered part of the main text.
% ---- Main text – up to 2,000 words, excluding the introductory paragraph, online Methods, references and figure legends. 
% ----References – as a guideline, we typically recommend up to 30.
% ----Display items – 3-4 items (figures and/or tables). 

% #################################################################
% #################################################################
% This is the suggested author list so-far. NOT FINAL! /JES ##########
% #################################################################
% #################################################################

\author{Joachim E. Sestoft}
\affiliation{Center for Quantum Devices \& Nano-science Center, Niels Bohr Institute, University of Copenhagen, 2100 Copenhagen, Denmark}
\affiliation{These authors contributed equally}

\author{Thomas K. Jensen}
\affiliation{Division of Synchrotron Radiation Research, Department of Physics, and NanoLund, Lund University, Box 118, Lund 221 00, Sweden}
\affiliation{These authors contributed equally}

\author{Vidar Flodgren}
\affiliation{Division of Synchrotron Radiation Research, Department of Physics, and NanoLund, Lund University, Box 118, Lund 221 00, Sweden}

\author{Abhijit Das}
\affiliation{Division of Synchrotron Radiation Research, Department of Physics, and NanoLund, Lund University, Box 118, Lund 221 00, Sweden}

% \author{Nathanael Löfström}
% \affiliation{Division of Synchrotron Radiation Research, Department of Physics, and NanoLund, Lund University, Box 118, Lund 221 00, Sweden}

\author{Rasmus D. Schlosser}
\affiliation{Center for Quantum Devices \& Nano-science Center, Niels Bohr Institute, University of Copenhagen, 2100 Copenhagen, Denmark}

\author{David Alcer}
\affiliation{Division of Solid State Physics, Department of Physics, and NanoLund, Lund University, Box 118, Lund 221 00, Sweden}

\author{Mariia Lamers}
\affiliation{Division of Solid State Physics, Department of Physics, and NanoLund, Lund University, Box 118, Lund 221 00, Sweden}
\affiliation{Wallenberg Initiative Materials Science for Sustainability, Department of Solid state physics, Lund University, 221 00 Lund, Sweden.}

\author{Thomas Kanne}
\affiliation{Center for Quantum Devices \& Nano-science Center, Niels Bohr Institute, University of Copenhagen, 2100 Copenhagen, Denmark}

\author{Magnus T. Borgström}
\affiliation{Division of Solid State Physics, Department of Physics, and NanoLund, Lund University, Box 118, Lund 221 00, Sweden}
\affiliation{Wallenberg Initiative Materials Science for Sustainability, Department of Solid state physics, Lund University, 221 00 Lund, Sweden.}

\author{Jesper Nyg\aa rd}
\affiliation{Center for Quantum Devices \& Nano-science Center, Niels Bohr Institute, University of Copenhagen, 2100 Copenhagen, Denmark}
%\email{nygard@nbi.ku.dk}

\author{Anders Mikkelsen}
\affiliation{Division of Synchrotron Radiation Research, Department of Physics, and NanoLund, Lund University, Box 118, Lund 221 00, Sweden}

\begin{abstract}
%Abstract word count: 153. 

Computational hardware designed to mimic biological neural networks holds the promise to resolve the drastically growing global energy demand of artificial intelligence.\cite{mehonic2022brain, kudithipudi2025neuromorphic} A wide variety of hardware concepts have been proposed\cite{tsakyridis2024photonic, schuman2022opportunities,sze2017hardware}, and among these, photonic approaches offer immense strengths in terms of power efficiency, speed and synaptic connectivity. However, existing solutions have large circuit footprints\cite{markovic2020physics, zhang2020neuro} limiting scaling potential and they miss key biological functions, like inhibition.\cite{hu2024electronically} We demonstrate an artificial nano-optoelectronic neuron with a circuit footprint size reduced by at least a factor of 100 compared to existing technologies\cite{winge2020implementing, winge2023artificial, tsakyridis2024photonic} and operating powers in the picowatt regime. The neuron can deterministically receive both exciting and inhibiting signals that can be summed and treated with a non-linear function. It demonstrates several biological relevant responses and memory timescales, as well as weighting of input channels. The neuron is compatible with commercial silicon technology, operates at multiple wavelengths and can be used for both computing and optical sensing. This work paves the way for two important research paths: photonic neuromorphic computing with nanosized footprints and low power consumption, and adaptive optical sensing, using the same architecture as a compact, modular front end.

\end{abstract}
%\subsection*{Introduction}

%% Dummy intro
%The rapid evolution of neuromorphic computing has fueled the demand for innovative approaches that seamlessly merge diverse materials, improved operation speeds, reduced power consumption and circuit footprint.
\maketitle

The rising energy demand of artificial intelligence infrastructure is not sustainable.\cite{mehonic2022brain} Neuromorphic hardware offers encouraging solutions to this problem by mimicking the energy-efficient biological brain.\cite{kudithipudi2025neuromorphic, schuman2022opportunities} Many different hardware solutions have been suggested, but photonic components are especially promising in terms of speed and power-efficiency\cite{markovic2020physics, zhang2020neuro} and similar computing hardware can also serve as optical signaling/sensory systems. However, many still lack one or more essentials,\cite{markovic2020physics, zhang2020neuro, hu2024electronically, tsakyridis2024photonic} like: (i) miniaturized building blocks for high density integration; (ii) excitation and inhibition in the same device; (iii) linear fan-in (summation) and tunable nonlinear activation; (iv) low optical energy per operation; (v) controlled device-to-device variation and simple tunable weighting; (vi) wavelength selectivity for routing; and (vii) CMOS-compatible materials and processing, with a clear path to all-optical and on-chip links. For optical sensory systems many of the same demands are highly relevant in order to mimick the exceptional analytical power of the biological retina. This includes contrast resolution over many orders of magnitude of background light, excellent dynamic range and edge resolution. For intensity adaptation and edge sharpening inhibition plays an important role.\cite{miller2005physiology, mazade2016light, franke2017general}

%Existing photonic approaches are often area inefficient, which limits scaling, and many omit inhibition.\cite{markovic2020physics, zhang2020neuro, hu2024electronically, tsakyridis2024photonic} 

Here we combine three semiconductor nanowires to construct an artificial \textit{optical/electronic} (O/E) neuron, that fulfills these requirements. The active area is 30-90 µm\textsuperscript{2} (at least 100 times smaller than prior on-chip photonic activators)\cite{tsakyridis2024photonic, winge2020implementing,winge2023artificial} it provides both excitation and inhibition, sums concurrent optical inputs, and provides sigmoid activation functionality. It operates at pico-watt optical powers, shows millisecond-scale responses with 0.1-1 s recovery, can potentially support ~1 GHz operation,\cite{winge2023artificial} is responsive across multiple wavelengths, and offers voltage–tunable sensitivity/weighting. Compared with planar photonic platforms, our nanowire node provides an exceptionally high absorption cross-section per footprint together with bandgap-, geometry- and orientation-controlled wavelength and polarization selectivity. Nanowire technologies are highly refined and employed in many different technology areas (e.g. quantum computing\cite{badawy2024electronic} and solar cells\cite{otnes2017towards}) and the nodal architecture allows for combinations of diverse functionalities in a modular fashion, while being CMOS-compatible.\cite{maartensson2004epitaxial, jia2019nanowire, mauthe2020high}

%%%%%%%%%%%%% Fig. 1 %%%%%%%%%%%%%%%%%%%%%
\begin{figure*}[ht]
\vspace{0.2cm}
\includegraphics[scale=0.98]{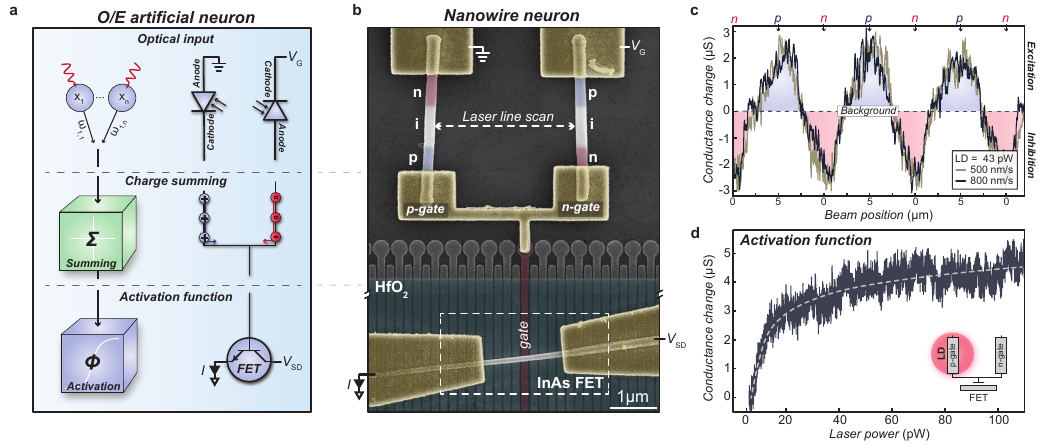}
\vspace{-0.10cm}
\caption{\textbf{Optoelectronic nanowire neuron.} \textbf{a}, Conceptual artificial \textit{optical/electronic} neuron and its circuit diagram. Optical inputs are converted to electrical charge carriers by two photodiodes. The carriers are summed and their total electrostatic signal is forwarded to a non-linear activation component (FET). \textbf{b}, False-coloured electron micrograph of the optoelectronic nanowire neuron device consisting of two InP pin-diode nanowires in opposing polarity connected electrostatically to an InAs nanowire-based field effect transistor. Yellow, Ti/Au contacts; Blue, p-doped region; Red, n-doped region; Grey, intrinsic InP or InAs; Dark red, Ti/Au gate; Light-blue, HfO\textsubscript{2}. Electrical setup is composed of source-drain bias, $V_\mathrm{SD}$, measured current, $I$, and applied voltage across the InP photodiode nanowires, $V_{G}$. Note: Image is cropped for illustrative purposes indicated by the two 'double slashes'. \textbf{c}, Conductance change as a function of laser spot position oscillating between the two photodiodes along the dashed line in \textbf{b}. Horizontal dashed line corresponds to the background conductance. \textbf{d}, Conductance change versus laser power indicating a sigmoid-like trace. The power meter at present setup calibrations does not detect linearly for power input \textless 6 pW. Inset shows the stationary beam spot position and grey dashed line serves as a guide to to the eye. }
\label{fig1}
\vspace*{-3mm}
\end{figure*}

\subsection*{Architecture, spatial mapping and nonlinear activation}

The functionality and circuit diagram of the neuron are illustrated in Fig. \ref{fig1} \textbf{a}. As excitatory and inhibitory optical receivers, we use two nanowire photodiodes and short their anode and cathode together by a metallic lead. The electron-hole pairs generated by the photodiodes are summed on this lead and provide charge to a gate affecting the electronic conductance of a nanowire field effect transistor (FET). In Fig. \ref{fig1} \textbf{b} we show an electron micrograph of a measured neuron device. We use InP nanowires with highly doped p-i-n junctions as photodiodes and connect the p-doped and n-doped regions of the two nanowires by Ti/Au  leads. The lead is extended to a predefined Ti/Au lead partially covered by HfO\textsubscript{2} serving as a high-$\mathrm{\kappa}$ dielectric insulator. It is electrically isolated from an intrinsic InAs nanowire contacted in the same processing step as the InP nanowires. We denote the two InP nanowires 'p-gate' and 'n-gate' depending on which doping polarity is connected to the gate. All nanowires are deposited deterministically using a micro-manipulator needle.\cite{flodgren2025flexible} To probe the nanowire-based neuron devices, we use (1) optical-beam-induced current (OBIC) on the photodiodes and (2) ac lock-in conductance measurements on the FET. The OBIC setup comprises a piezo stage and an optical microscope receiving a single mode laser fiber (663 nm) and uses a 100x objective lens to achieve a beam spot size of approximately 0.8 µm. This setup allows us to record time-resolved conductance modulations across the transistor nanowires as a function of optical beam position and illumination power. See Methods and Supplementary Information, S1, for details on device operation and measurement setup.

To demonstrate neural excitation and inhibition of the nanowire neuron, we first probe the device in terms of space- and optical power-resolved measurements using a single light source. In Fig. \ref{fig1} \textbf{c}, we plot the conductance change ($\Delta G$) across the nanowire FET as a function of beam spot position, as indicated by the dashed line in \textbf{b}. The laser diode is kept at a constant illumination power of 43 pW, placing the device in the saturated regime (see Fig. \ref{fig1}\textbf{d}). A voltage (\textit{V}$_{\mathrm{G}}$ = -3 V) is applied to the n-gate InP nanowire bringing both InP diodes into reverse bias, improving photosensitivity. The beam spot is moved back and forth between the two pin-doped nanowires in oscillatory motion. Here we observe an increase in conductance (excitation) when the laser spot approaches the p-gate and a decrease (inhibition) when approaching the n-gate. The horizontal dashed line corresponds to the background conductance measured in dark conditions. We attribute this behavior to the charge produced by the photodiode nanowires, which alters the conductance of the InAs FET through electrostatic gating, where the sign of this modulation is determined by the doping polarity of the connected end of the nanowire. We control for reproducibility by performing these measurements for two distinct piezo-stage-stepping speeds. The traces overlap, however, faster operation may introduce hysteresis. These measurements demonstrate optically-controlled conductance modulations akin to biological neural excitation and inhibition, and mimics light signal based synaptic communication originating from multiple on-chip or off-chip sources.

%($\textit{n}\rightarrow\textit{p}\rightarrow\textit{n}\rightarrow$ ...)

Next, we fix the beam spot position on the excitatory p-gate nanowire and record the change in conductance versus optical power (Fig. \ref{fig1} \textbf{d}). For lower ranges of input power ($\sim$6-12 pW) the device exhibits a significant quasi-linear conductance increase that saturates for lasing powers above $\sim$60 pW. The sigmoid-like dependence of the device conductance versus optical power is highly suitable for neural thresholding and activation functionality, providing the crucial nonlinear part in neural networks.\cite{stone2017anatomically} As discussed later weighting of individual optoelectronic nanowire neurons can be achieved by tuning their optical sensitivity using specific $V_{\mathrm{G}}$ values modulating their activation curves. 

%The inserted panel shows the zoom-in around the lower limit of detection of the input power measurements.
%Balancing the weights of individual nodes in a neural network is key to achieving network learning.

\begin{figure}[tb!] 
\vspace{0.2cm}
\includegraphics[scale=0.95]{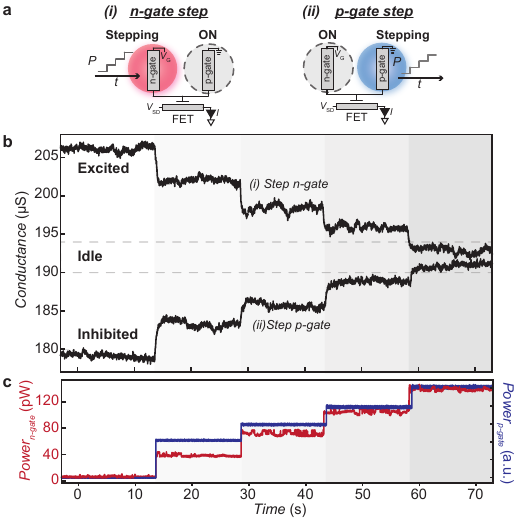}
\vspace{-0.10cm}
\caption{\textbf{Summing optical inputs.} \textbf{a}, Measurement setup configuration where \textbf{\textit{(i)}} denotes incident optical power stepped on the n-gate with constant light on the p-gate (ON) and \textbf{\textit{(ii)}} shows the inverted configuration. \textbf{b}, Conductance across the FET plotted as a function of time. Traces are correlated with the two schematics shown in panel \textbf{a}. Idle state is defined as the background conductance ($\pm$2 µS) of the device under dark conditions, where $V_{\mathrm{G}} = -3$ V. \textbf{c}, Power stepped of the light sources directed to the n-gate and p-gate in four increments as a function of time.}
\label{fig2_2}
\vspace*{-3mm}
\end{figure}

%\textbf{c}, Power stepped of the light emitting diode (LED) in four increments as a function of time.
%Purple trace corresponds to constant LD light directed to the n-gate while the power of LED illuminating the n-gate is stepped as indicated in \textbf{c}.

%Maybe we find a way to state that 'ídle state' is a somewhat arbitrary definition, but that the neuron behavior for sure is qualitatively sound, and almost quantitatively also:)

%and exposing it to constant LED illumination. 

%applying an \textit{optical iris} to 
% in steps of $\Delta P_{\mathrm{LD}}$ $\sim 35$ pW

%. This results in a distinct power-dependent reduction of the net positive charge on the electrostatic gate affecting the FET conductance, transitioning the neuron device from its excited state back to the idle state, 
%The purple trace corresponds to the 'inverse' experiment, serving as the control experiment. Here we expose the n-gate to constant LD illumination, now bringing the device into an inhibited state before recording the conductance and stepping the LED directed to the p-gate (Fig. \ref{fig2_2}\textbf{c}). From its inhibited state we observe how the neuron device is brought back to its idle state as the p-gate receives increasing optical power. 
%-- that is, summation of positive and negative charge carriers.
%across the FET
\subsection*{Summing multiple optical inputs}

To investigate optical input summation, we perform measurements on both the n- and p-gates simultaneously. First, the neuron device is transitioned to an excited/inhibited state by selectively illuminating either the p- or n-gate. These states are arbitrarily defined as neuron conductance $\sim$10 µS above/below its conductance in dark conditions (idle). Next, we increase the optical power incrementally on the other, so-far un-illuminated gate, using another light source. Figure \ref{fig2_2}\textbf{a} and \textbf{c} show the measurement configurations and applied powers, respectively. In Fig. \ref{fig2_2}\textbf{b} we show the simultaneously recorded conductance. Here we observe two traces with distinct step-like conductance modulations as the optical power is incrementally increased on either the n-gate (\textbf{i}) or the p-gate (\textbf{ii}). As either gate receives increasing optical power, the rate of charge carriers produced by the photodiode nanowire increases. These carriers recombine across the metal gate with the charge carriers (of opposite sign) being generated by the light incident on the other photodiode. Hence balancing the rates of the different carriers being generated, brings the device from its excited/inhibited state back to its idle state in a step-like manner. This result demonstrates the ability of the neuron device to sum two or more independent optical signals, a critical component enabling fanning in signals from other nodes. 

%\cite{flodgren2024} 

\subsection*{Temporal dynamics and memory}

\begin{figure*}[tb!]
\vspace{0.2cm}
\includegraphics[scale=0.98]{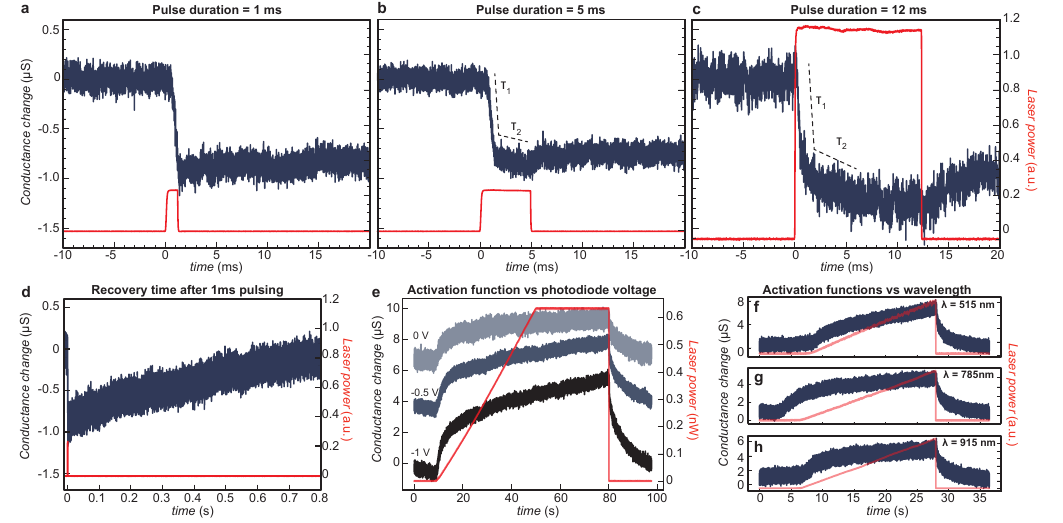}
\vspace{-0.10cm}
\caption{\textbf{Time dynamics, activation function tunability and wavelength sensitivity.} \textbf{a-c}, Conductance change versus time recorded during three light pulse sequences of 1, 5 and 12 ms and different intensities. The light is directed onto the n-gate to generate 'inhibitory' behavior. Panels \textbf{a-c} show initial behavior during pulsing whereas \textbf{d} shows the time to recover to baseline conductance. During these measurements the p-gate is kept under constant selective illumination at power ranges of 10-20 mW. \textbf{e}, Conductance change recorded as a function of time and laser power for three different photodiode voltages, $V_{\mathrm{G}} =$ 0, -0.5 and -1 V. As $V_{\mathrm{G}}$ is tuned more negative the magnitude of the sigmoidal activation function shape becomes more pronounced. Traces are offset for clarity. \textbf{f-h}, Conductance change versus time and laser power recorded using three different wavelengths, $\lambda =$ 515, 785 and 915 nm. All data are event-averaged.}
\label{fig3}
\vspace*{-3mm}
\end{figure*}

Following, we explore the temporally-resolved dynamics of the nanowire neuron. In Fig. \ref{fig3}\textbf{a-c} we show conductance change recorded as a function of time and laser power. During these measurements light pulses of varying duration (1, 5 and 12 ms) and intensities are directed onto the n-gate resulting in inhibitory pulsing behavior. Similar measurements on the p-gate showing excitatory behavior are presented in Supplementary Information S2. In these traces we observe two different time dependent behaviors. For all pulse durations and intensities a fast decrease in conductance is observed within $\sim 1$ ms with a $\Delta G \sim -1$ µS. During pulses longer than 1 ms the conductance continues to decrease albeit at a slower rate before it saturates. This behavior is highlighted by the dashed lines in Fig. \ref{fig3}\textbf{b-c}. Comparing the slopes of the slower timescales ($\tau_2$), we find that as intensity increases the negative slope of $\tau_2$ increases too. Additional recordings on variable and constant spiking rates are shown in Supplementary Information S3.

%\textit{Should we guess what is happening here specifically?}
%%%%%%%%%%%%% Fig. 3 %%%%%%%%%%%%%%%%%%%%%

Next, we explore the timescales required for the neuron device to reset to its idle state (memory properties). In Fig. \ref{fig3}\textbf{d} we show a recording of the time extending after the 1 ms pulsing measurement from Fig. \ref{fig3}\textbf{a}. Here we observe a sharp decrease in conductance before conductance returns to baseline after about 0.8 s. In other measurements we find the reset time to be approximately $100$ ms. In current device geometries the operation speeds of the neuron devices are biologically relevant with millisecond reaction speeds and memory timescales on the order of 100 milliseconds to several seconds.\cite{zucker2002short} We attribute the reaction times of the neuron devices primarily to the response times in the FET component. Factors such as capacitances, surface states, traps, gate dielectrics, and long channel lengths are known to control response times in transistors based on III/V materials and can be engineered for desired functionalities.\cite{del2011nanometre} In addition, the response times could be influenced by additional factors, such as running the FET outside of its linear operation. When operating the neuron at 20 pW (for a $\sim 1$ µS conductance change) we estimate the energy use to approx. 200 fJ per operation (see Supplementary information, S4, for further details).

% of achieving tunable memories into the nanowire neurons.

%However, high-speed InAs nanowire-based FETs can be routinely operated in the GHz range, and there are no fundamental obstacles in integrating these types of nanowire FETs into the nanowire neuron. 

Tuning the activation function of nodes in a neural network is a crucial task. In Fig. \ref{fig3}\textbf{e} we show three different activation functions measured under different $V_{\mathrm{G}}$ applied across the two nanowire photodiodes. As $V_{\mathrm{G}}$ is tuned to larger negative values the magnitude of the non-linear response of the device becomes more pronounced. This is consistent with bringing the two InP-photodiodes deeper into reverse bias and increasing sensitivity due to widening of the intrinsic channel enhancing light absorption and limiting intrinsic carrier recombination. Additional information is presented in the Supplementary Information S5. Relying on this approach individual nodes can be switched ON and OFF while also tuning their activation functions dynamically. Lastly, we demonstrate how the neuron can receive optical input of different wavelengths and generate activation functions as long as the energy of the incident light is larger than the bandgap of the nanowire photodiode material. Here we observe similar non-linear activation functionality, all with sigmoidal-shapes, as shown in Fig. \ref{fig1}\textbf{d}. We show three examples using wavelengths of $\lambda$ = 515 nm, 785 nm and 915 nm. In the Supporting Information S6 we show how below-band gap optical pulses leave the neuron device unaffected. Hence selecting material compositions with desired energy bandgaps provides the neurons with tunable wavelength sensitivity enabling optical communication between only select nodes, or designing distinct excitatory/inhibitory neural pathways in the same nodes.

\subsection*{Discussion}

%%% V1 of the discussion. %%%%

Substituting typical optical components such as waveguides, ring modulators and photodetectors ($\sim$100 µm to 1 mm scale) with nanoscaled optoelectronic crystals is a promising route to significantly reducing circuit footprints for optical information processing. The active on-chip area of the nanowire-based neuron device (see details in the Supplementary information, S7) ranges from 30-90 µm\textsuperscript{2}. Comparing to state-of-the-art optical activation function demonstrations\cite{tsakyridis2024photonic} this provides a footprint reduction between two to eight orders of magnitude depending on the compared platform. In addition, our neuron can host more synaptic connections by integrating additional inhibiting/exciting wires of different material or polarization sensitivity, and can also naturally receive multiple signals from surrounding wires in a broadcasting scheme, even in 3D.\cite{draguns2025}

Further footprint reduction and scalable production can be achieved by advanced nanowire placement techniques.\cite{jia2019nanowire} For large ensembles of in-plane neural nodes, the fibre-like morphology of nanowires, resembling biological axons, could emulate high-density connectivity, providing further miniaturization and recent work has shown direct on-chip optical communication between single nanowires providing synaptic capabilities.\cite{flodgren2024} Moving towards 3-dimensional interconnects between nanowire neurons relying on vertical nanowire arrays pioneered in the high-speed nanowire-based FET community\cite{bryllert2006vertical} would enable further miniaturization. In neuromorphic systems, analog-to-digital conversion blocks put high demands on both power and area consumption. In this context, the modular nature of our nanowire neuron holds promise as the interface of analog-to-digital converters to other optical neural networks for read-out and digitization.

Most existing nanowire-based neural network approaches rely on randomly dispersed nanowires mostly using memristive dynamics.\cite{kuncic2021neuromorphic} Hence, in contrast this work provides a significantly different approach to nanowire-based neural networks, yielding a generalizable nodal architecture with highly optimized area efficiency while providing advantages of optical neural networks like speed and power consumption. Critically, the field of nanowire technology is highly mature and diverse, offering many routes for integration of desired functionality into novel nanowire-based neuromorphics while being compatible with CMOS technologies.\cite{maartensson2004epitaxial, jia2019nanowire, mauthe2020high} 

%This might be a key ingredient for the success of these emerging neuromorphic platforms.\cite{kudithipudi2025neuromorphic} 

% capable of acting as building blocks in larger networks
%, our nanowire neuron is assembled deterministically as opposed to , they do not provide , and previous approaches are not light-based. In addition, these random network architectures do not provide the same critical miniaturization routes as the deterministic nanowire neuron.

%.\cite{} Randomly distributed nanowire networks have shown great promise, however,

%%%%%%%%%%%%%%%%%%%%%%%%%%
%%% Maybe the following goes at the end of the discussion of Figure 2? And is this discussion justified at all? %%%

% In neuromorphic systems, analog-to-digital conversion blocks put high demands on both power and area consumption. In this context, the modular nature of our nanowire neuron holds promise as the interface of analog-to-digital converters to optical neural networks for read-out and digitization. 
% What could the bit depth be? Judge from signal-to-noise ratio? This depth can be extended dramatically by improving device engineering.   

%%%%%%%%%%%

Several future studies related to this work are important. (1) Using GaInP for the inhibitory photodiode could create a built-in spectral window: photons just above the InP band edge would activate the excitatory channel, whereas higher-energy photons above the larger GaInP edge would also engage inhibition and may be suppressed. Tuning the Ga fraction could set the upper cutoff, while the excitatory material fixes the lower. (2) Supplying controlled background illumination to the inhibitory input might let each pixel report local contrast rather than absolute intensity, enabling retina-like adaptation over a wide dynamic range without extra circuitry. (3) If each inhibitory input receives a weighted sum from neighboring pixels, via shared optics or simple interconnects, the array could implement center–surround filtering that enhances centers, suppresses backgrounds, and potentially sharpens edges and reduces noise before downstream processing. (4) Integrating an optical-output\cite{flodgren2024} to the nanowire neuron establishes an O/E/O building block that, when integrated into interconnected arrays, supports all-optical neural computations.\cite{winge2020implementing, winge2023artificial} To achieve this, engineering tasks such as optimization of the FET geometry, threshold voltages and ON/OFF ratios, will be critical for the neuron device to control and power a nanowire-based LED optical output, as the neuron device in its current state provides a modulation of conductance of 5-12$\%$. We note that an optimized O/E/O building block should run at operation speeds in the 1 GHz regime,\cite{winge2023artificial} and use an estimated power consumption of 10~nW per node.\cite{winge2020implementing} To tune the memory of the neuron devices we predict several promising routes. For one, nano-floating-gate structures\cite{ansari2021core,lee2021non} can be used to engineer leaky or non-volatile memories in nanowire FETs. Other approaches entail engineering of synaptic memory like adding switchable molecular dyes between nanowire neurons.\cite{alcer2025integrating} 

In conclusion, we provide a nanowire-based and deterministic platform capable of drastically reducing the circuit footprint of all-optical neural networks and next-generation adaptive optical sensors. These are expected to be especially relevant for lightweight edge neural networks where direct coupling to optical inputs allow for on-device, AI-enhanced sensing.

\subsection*{Methods}

\textbf{InP nanowire growth.} The InP nanowires were synthesized via metalorganic vapor phase epitaxy (MOVPE) on patterned substrates. Gold seed particles, defined by nanoimprint lithography into hexagonal arrays with a pitch of 0.50 µm, served as catalytic growth sites. Growth proceeded in a laminar flow MOVPE reactor (Aixtron 200/4), operating at 100 mbar total pressure using hydrogen (H\textsubscript{2}) as carrier gas at a flow rate of 13 L/min. Prior to growth, substrates underwent a preanneal nucleation step at 280 $^{\circ}$C, including TMIn precursors and PH3.
This was followed by an annealing step at 550 $^{\circ}$C under a phosphine (PH\textsubscript{3})/H\textsubscript{2} ambient to ensure pattern integrity.\cite{otnes2016strategies} Subsequently, the reactor temperature was lowered to the growth temperature of 440 $^{\circ}$C. Growth was initiated by introducing trimethylindium (TMIn) and hydrogen chloride (HCl) into the gas stream. The doping profile consisted of a heavily controlled gradient: the bottom segment was uniformly n-doped using tetraethyltin (TESn) at \textsubscript{$\chi$}TESn = 4.3$\times$10\textsuperscript{-5}. The middle segment was either nominally intrinsic or lightly p-doped, with DEZn concentrations ranging between \textsubscript{$\chi$}DEZn = 0 and 2.1$\times$10\textsuperscript{-7}. Finally, the top segment was p-doped, with molar fraction varied from \textsubscript{$\chi$}DEZn = 0.09$\times$10\textsuperscript{-5} to 8.24$\times$10\textsuperscript{-5}.

\textbf{InAs nanowire growth.} A molecular beam epitaxy (MBE) system is used to grow Au-seeded wurtzite InAs nanowires, along the [0001]B direction on InAs (111)B substrates using the vapour-liquid-solid mechanism. Arrays of Au catalyst particles are placed via standard EBL with particle radius $r_{\mathrm{Au}} =$ 20-120 nm and height $h_{\mathrm{Au}} =$ 10-50 nm. After substrate annealing at As overpressure at $T = $ 500 $^{\circ}$C for 5 min, predominantly vertical nanowire growth is initiated at growth temperatures ranging from $T_{\mathrm{growth}} =$ 445-450 $^{\circ}$C. Axial nanowire growth is carried out for a duration of 10-120 min before a short break (5 min) is introduced and the As$_{\mathrm{4}}$/As$_{\mathrm{2}}$ ratio is increased.

\textbf{Device fabrication.} All devices are fabricated on highly doped Si$^{++}$ substrates covered by 200 nm of thermal oxide. Nanowires were picked up and placed semi-automatically using a tungsten-based micromanipulator needle under a 100x objective lens. Metallic leads to the nanowires were fabricated by electron beam lithography. Metallic leads were patterned using electron beam lithography, after which RF ion (Ar+) milling was performed in a metal deposition chamber, immediately followed by the e-beam deposition of Ti and Au (5 nm/300 nm) to create ohmic contacts to the nanowires. The bottom gates (5/30 nm) were defined similarly and covered by 20 nm of HfO\textsubscript{2} grown by atomic layer deposition.

\textbf{Measurements.} Optical-beam-induced current measurements were performed to spatially characterize the nanowire neuron devices. A continuous-wave laser diode emitting at 663 nm was focused near its diffraction limit (about 800 nm full width at half-maximum) onto the sample using a 100x objective lens. Precise spatial positioning of the spot relative to the nanowire device was achieved using a piezoelectric motor stage capable of sub-nanometer precision. As the laser beam was raster-scanned across the device, the generated photovoltage affecting the conductance of the InAs nanowire via the electrostatic gates was mapped (see details in Supplementary Information S1). Conductance \textit{g}=\textit{d}I/\textit{d}V\textsubscript{SD} was measured across the InAs nanowire component using a.c.-lock-in techniques with an excitation voltage in the range 20 µV–5 mV and integration times of 10 µs. 

% ac excitation voltages from 100 µV to 2 mV and frequencies of up to 100 kHz

\subsection*{Acknowledgements}

This work was supported by the Swedish Research Council, NanoLund, supported by Myfab, the Wallenberg Initiative Materials Science for Sustainability (WISE) funded by the Knut and Alice Wallenberg Foundation, Danish National Research Foundation (DNRF101), the Olle Engkvist Foundation, the Novo Nordisk Foundation project SolidQ, and the European Union Horizon Europe project InsectNeuroNano (Grant 101046790).

\subsection*{Author contributions}
%%%%%%%%% DRAFT %%%%%%%%%%%% Not final or set in stone. Please change accordingly. 
%%%%%%%%%%%%%%%%%%%%%%%%%%%%%
%%%%%%%%%%%%%%%%%%%%%%%%%%%%%

Conceptualization, J.E.S., J.N. and A.M.; Device fabrication and recipe development, J.E.S., V.F., A.D., and R.D.S.; Material development, D.A., T.K., M.L., M.T.B. and J.N.; Measurements, T.K.J and J.E.S.; Writing, J.E.S. and T.K.J.; Supervision, M.T.B., J.N. and A.M. 
%%%%%%%%%%%%%%%%%%%%%%%%%%%%%
%%%%%%%%%%%%%%%%%%%%%%%%%%%%%
%%%%%%%%%%%%%%%%%%%%%%%%%%%%%
\subsection*{Competing interests}
The authors declare no competing interests.

\subsection*{Supplementary Information} 

Supplementary Information is available for this paper. 

% Supplementary information containing extended information on  at \url{https://sid.erda.dk/sharelink/eeL2CxBB6p}.

\bibliography{ref}
\end{document}